 \documentstyle[12pt]{article}
\begin{document}
\title{Explicitly solvable cases of one-dimensional quantum chaos}
\author{R. Bl\"umel, Y. Dabaghian, and R. V. Jensen\cr
        Department of Physics, Wesleyan University,\cr
        Middletown, CT 06459-0155}
\date{\today}
\maketitle

\begin{abstract}
We identify a set of
quantum graphs
with unique and precisely defined spectral properties called
{\it regular quantum graphs}.
Although chaotic in their classical limit with positive
topological entropy,
regular quantum graphs are
explicitly solvable.
The proof is constructive:
we present
exact periodic orbit
expansions for individual energy levels, thus
obtaining an analytical
solution for the spectrum of regular quantum graphs
that is complete, explicit and exact.
\end{abstract}


%%%%%%%%%%%%%%%%%%%%%%%%%%%%%%%%%%%%%%%%%%%%%%%%%%%
%\begin{multicols}{2}

A quantum graph
\cite{NV1,NV2,NV3,Roth,QGT1,QGT2,QGT3,QGT4}
is a
network of vertices and bonds with a quantum particle
moving along its bonds. An example of a graph with
five vertices and six bonds is shown in Fig.~1.
The wave function and the
energy levels of a quantum particle on a graph are defined by the
corresponding one-dimensional Schr\"odinger equation.
Despite the apparent simplicity of the
system, quantum graphs have proven to be a rich source of
physical insight. From the mathematical point of view, the
spectral properties of
Schr\"odinger operators on graphs are highly
nontrivial and have been widely investigated in
the mathematical literature \cite{NV1,NV2,NV3,Roth}.
Quantum graphs and networks have also been used to model various
phenomena in different branches of
physics and chemistry for more than 30
years. The most recent physical
development appeared in a series of
publications \cite{QGT1,QGT2,QGT3,QGT4},
where quantum graphs were studied in
the context of quantum chaos.

It is easy to see that
the behavior of a
particle on a quantum graph is very complex.
Each time the particle encounters
a vertex
$v_{i}$ of the graph, it can
scatter with different probabilities
in the forward or backward directions along any of the
bonds emanating from the vertex.
As a simple physical
analogue of this system one may imagine a
beam of light traveling along a
network of optical fibers. At every joint
of the fibers the light waves
scatter in such a way that the total
energy flux is conserved.

As a result of the multiple scattering
possibilities at the vertices,
the dynamics of a classical particle
on a graph is very complex and
the number of possible periodic orbits
traced by the particle
increases exponentially with their lengths.
Consequently the topological entropy of the particle
is positive and
since the
phase space of the system is bounded,
the dynamics of the particle is
ergodic.
The classical chaoticity notwithstanding it was shown
that several important spectral
characteristics of quantum graphs,
such as the density of states and
the spectral staircase,
can be obtained exactly in terms of periodic
orbit expansion series
\cite{QGT1,QGT2,QGT3}.

The ``wiring'' of a quantum graph, i.e. the
arrangement of bonds and vertices, is called
the topology of the quantum graph. For any given graph
topology there exists a wide variety of possible
quantum graphs.
The vertices, for instance, may be realized as
simple hubs that redistribute the quantum flux into
various channels; or we may place
$\delta$ function potentials
of various strengths at the vertices; or there can even be
potential functions on the bonds. Because of this
flexibility we anticipate that the topology of
a quantum graph alone does not uniquely specify
the graph's  spectral properties. We conjecture that there
are several different types of
quantum graphs, all chaotic in their classical limit,
but each exhibiting unique and precisely defined
spectral characteristics.

We start a rigorous classification of quantum graphs
by defining {\it regular quantum graphs}.
Although generically their classical limit is chaotic,
we show that
their spectrum is explicitly solvable
analytically, state by state,
via explicit periodic orbit expansions.
This result is backed up by rigorous mathematical
proofs \cite{Kar} whose basic elements are presented
below.
To the authors' knowledge this is the first time that
the spectrum of
a quantum chaotic system is obtained
{\it exactly} and {\it explicitly}.
In addition
regular quantum graphs show a spectral gap at small
energy spacings and a cut-off at large spacings.
Both the size of the gap and the location of the cut-off
are computed analytically.

The regular quantum graphs are different from a class of
quantum graphs described in the literature
\cite{QGT1,QGT2,QGT3} whose
level spacing distributions, to a good approximation,
exhibit features of the Gaussian orthogonal (unitary)
ensemble
\cite{BGS,LH}.
Thus there are at least two different types of
quantum graphs with distinct spectral properties.

The periodic
orbits of graphs are defined as the
periodic connected sequences
of bonds $B_{ij}$. Denoting by $k$ the wave number (momentum)
of the particle, each periodic orbit
contributes
\begin{equation}
S_{ij}=\int_{B_{ij}}k(x)\, dx,  \label{action}
\end{equation}
to the total action of a path traced by the particle.
It turns out that the information contained
in the totality of all the
possible classical periodic orbits often allows one to
reconstruct {\it exactly}
certain quantities of a purely quantum nature.
For example, according to recent results
\cite{Roth,QGT1,QGT2,QGT3}, the exact periodic
orbit expansion for the density of states
can be written as
\begin{equation}
\rho (k)\equiv \sum_{j=1}^{\infty }\delta \left(k-k_{j}\right) =
\bar{\rho}(k)+\frac{1}{\pi }\mathop{\rm Re}
\sum_{p}T_{p}\sum_{\nu=1}^{\infty}A_{p}^{\nu }\,e^{i\nu S_{p}(k)}.
\label{ro}
\end{equation}
Here $\bar{\rho}(k)$ is the
average density of states, $\nu$ is the repetition index,
$T_{p}=\partial S_p(k)/\partial k$, and $S_{p}$,
$A_{p}$ are correspondingly
the action and the weight factor of
the prime periodic orbit
labeled by $p$.
We assume
in what follows that the system is scaling
\cite{Kar,RS2,Koch,Bauch,Found,us,Tom}.
This means that the actions of the periodic orbits
are proportional to the wave number,
\begin{equation}
S_{p}(k)=S_{p}^{0}k,
\label{scale}
\end{equation}
where $S_{p}^{0}$, a constant,
is the reduced
action. In this case $T_{p}=S_p^0$ and
$A_{p}$ are $k$-independent constants.
We define the total reduced action $S_0$ of
the graph, $S_0=(\int k\, dx)/k$, where the integral
is extended over all of the bonds of the graph.
The scaling assumption
is not an artificial
restriction. It occurs, for instance, in atomic physics where
scaled spectroscopy is now a common experimental
technique \cite{Tom}. In addition,
scaling quantum systems of this kind are the analogues of certain
electromagnetic ray-splitting
systems, flat metal cavities partially filled
with a dielectric substance such as Teflon
\cite{Koch,Bauch,Found}.

The formal description of a quantum graph system
proceeds as follows \cite{QGT1,QGT2,QGT3}. On
a bond $B_{ij}$ connecting the vertices $v_{i}$
and $v_{j}$, the wave
function of a quantum particle is defined by the
one-dimensional
Schr\"odinger equation which may include potentials
on the bonds \cite{us}.
At every vertex $v_{i}$ the wave function satisfies the usual
boundary
conditions of continuity and flux conservation.
The consistency of the boundary
conditions at every vertex of the graph
naturally yields the global quantization
conditions that determine the
momentum eigenvalues $k_{n}$.
A simple and elegant method based on the
scattering quantization approach was
presented in \cite {QGT1,QGT2,QGT3}, where
the quantization condition is
given in the form
\begin{equation}
\det [1-S(k)]=0.
\label{det}
\end{equation}
Here $S(k)$ is the scattering
matrix \cite{QGT1}, written explicitly in terms
of the connectivity matrix \cite{QGT1} of the graph.
It can be shown that the modulus
of the complex function (\ref{det}) is a
trigonometric function of the form
\begin{equation}
\cos(S_0k+\pi\gamma)-\Phi(k) = 0,
\label{eqn}
\end{equation}
where
\begin{equation}
\Phi(k)=\sum_{i}a_{i}\cos(S_{i}k+\pi\gamma_{i}).
\label{Phi}
\end{equation}
In the scaling case
the $a_{i}$ are constants and $S_{i}< S_0$ are certain
combinations of the classical actions (\ref{action}).
In general the functions $\gamma$
and $\gamma_{i}$ are bounded and tend to
constant values for $k\rightarrow \infty$.
In the scaling case they are
$k$-independent constants.

In order to proceed we define
regular quantum graphs.
A regular quantum graph
fulfils the condition
\begin{equation}
\alpha=\sum_i\mid a_i|<1.
\label{regul}
\end{equation}
It is convenient for the following discussion to assume
$\alpha>0$, i.e. we exclude trivial graphs with
$\alpha=0$. They are regular quantum graphs whose spectrum can
be obtained trivially.
For regular quantum graphs (\ref{eqn}) can be solved
formally for the momentum eigenvalues
$k_{n}$.
We obtain the following implicit
solution for the roots of this quasi
periodic function
\begin{equation}
k_{n}={\pi\over S_0}\left[n+\mu-\gamma\right] + {1\over S_0}
\cases{\arccos(\Phi_n), &for $n+\mu$ even \cr
        \pi-\arccos(\Phi_n), &for $n+\mu$ odd, \cr}
\label{levels}
\end{equation}
where $\Phi_n=\Phi(k_n)$ and $\mu\in Z$, a fixed integer,
is to be chosen such that $k_1$ is
the first nonnegative solution of
(\ref{eqn}).
Because of (\ref{regul}),
the boundedness of the trigonometric functions
in (\ref{Phi}) and the properties of the $\arccos$ function,
the second term of (\ref{levels}) is bounded away from
0 and $\pi/S_0$ and assumes only values between
$u$ and $\pi/S_0-u$,
where $0<u=\arccos(\alpha)/S_0<\pi/2S_0$.
Thus, for regular graphs the points
\begin{equation}
\hat{k}_{n}=\frac{\pi }{S_0}(n+\mu-\gamma+1), \ \ n=1,2,\ldots
\label{k}
\end{equation}
are not roots of (\ref{det}). They serve as separators
between root number $n$ and root number $n+1$.
In fact (\ref{levels})
implies even more: the existence of finite-width
``root-free zones'' $R_n=(\hat k_n-u,
\hat k_n+u)$
surrounding every separating
point $\hat k_n$, where no roots of (\ref{det}) can be
found. Thus, roots of (\ref{det}) can only be found
in the ``root zones'' $Z_n=[\hat k_{n-1}+u,\hat k_n-u]$,
subsets of the root intervals
$I_n=[\hat k_{n-1},\hat k_n]$.
Since $S_0$ is the largest action in (\ref{eqn}) and
(\ref{Phi}), it can be shown \cite{Kar} that
$k_n$ is the {\it only} root in $Z_n$. Therefore,
in summary, there is exactly one root $k_n$ inside of
$Z_n\subset I_n$, and this root is bounded away from
the separating points $\hat k_{n-1}$ and $\hat k_n$ by
a finite amount $u$.

The spectral properties of regular graphs discussed above
allow us
to draw several important conclusions.
(i) Since there is exactly one root $k_n$ of (\ref{det})
in $I_n$, this
proves rigorously that
the number of roots of (\ref{det}) smaller than
$k$ grows like $\bar N(k)\sim S_0k/\pi$
(Weyl's law), while the second
term of (\ref{levels}) represents
the fluctuations around the average.
(ii) The existence of the root-free zones $R_n$
gives rise to a
spectral gap of finite width
$g=2u$ in the
nearest neighbor spacing distribution of regular
quantum graphs, i.e. no level spacings smaller than
$g$ are ever found.
This, for instance, precludes the possibility of
degenerate eigenvalues. (iii)
The existence of the
separating points (\ref{k})
together with the root-free zones
imply the existence of a spectral cut at
$c=2(\pi/S_0-u)$, i.e. no level spacings greater than
$c$ are ever found.
(iv)
The existence of the
separating points (\ref{k}) and the root-free zones
$R_n$ are the key for obtaining
an explicit and
exact periodic orbit expansion for every root of (\ref{det}).
Multiplying both sides
of (\ref{ro}) by $k$ and integrating from $\hat k_{n-1}$ to
$\hat k_{n}$ we obtain
\begin{eqnarray}
k_n = \hat k_n - {\pi\over 2 S_0} -
{1\over\pi}{\rm Re}\sum_p \sum_{\nu=1}^{\infty}A_p^{\nu}\,
{e^{i\nu S_p^0\hat k_n}\over \nu } \left\{
(1-e^{-i\nu\omega_p})\left(i\hat k_n-{1\over\nu S_p^0}\right)
+{i\pi\over S_0}e^{-i\nu\omega_p}\right\},
\label{kn}
\end{eqnarray}
where $\omega_{p}=\pi S_{p}^{0}/S_0$.
Since all the quantities on the right-hand side of (\ref{kn})
are known,
this formula provides an explicit
representation of the roots $k_n$ of the
spectral equation (\ref{det}) in
terms of the geometric characteristics of
the graph. To our knowledge, this
is the first time that the energy levels
of a chaotic system are expressed
explicitly in terms of a periodic orbit
expansion. Previously, explicit
formulae for individual energy levels were
known only for integrable systems.
In the context of periodic orbit
theory, the energy levels of integrable
systems are given by the Einstein-Brillouin-Keller
(EBK) formula \cite{EBK}.
However, apart from a few exceptional
cases \cite{Szabo} EBK quantization is
only of semiclassical accuracy.

For a generic chaotic system the energy
levels can only be obtained
indirectly as the singularities of the
periodic orbit expansion of the
density of states (Gutzwiller's
formula \cite{EBK}), an implicit method
which, in general, is only of semiclassical accuracy.
Formula (\ref{kn}), on the other
hand, shows that for regular
quantum graphs every
quantum level can be obtained
{\it individually}, {\it explicitly} and {\it
exactly} in terms of classical parameters.

In order to demonstrate that the class of regular
quantum graphs is not empty we present
an explicit example: the
one-dimensional step potential
shown in Fig.~2 (a)
\cite{Flugge}.
With
\begin{equation}
V_0=\lambda E
\label{scal}
\end{equation}
we turn it into a scaling system.
Physically this potential is realized
for a rectangular microwave cavity partially
loaded with a dielectric substance
\cite{Koch,Bauch,Found}.
The scaling step potential is
equivalent to the scaling
three-vertex chain graph shown in
Fig. 2 (b), with two bonds $L_{1}=b$ and $L_{2}=\beta(1-b)$,
$\beta=\sqrt{1-\lambda}$.
In this case the spectral equation is given by
\begin{equation}
\det\left[ 1-S(k)\right]=
\sin(Lk)-r\sin\left[\left(L_{1}-L_{2}\right)k\right] =0,
\label{3chain}
\end{equation}
where $L=L_{1}+L_{2}$, and $r=(1-\beta)/(1+\beta)$
is the reflection
coefficient at the vertex $v_{2}$ between the two bonds.
The solution of (\ref{3chain}) is
\begin{equation}
k_{n}=\frac{\pi n}{L}+(-1)^n\frac{1}{L}
\arcsin \left[r\sin(\left(L_{1}-L_{2}\right)
k_n)\right],
\label{rbound}
\end{equation}
where the second term, which gives the
fluctuations around the average, is
bounded by
\begin{equation}
\left| \delta k_{n}\right| \leq \frac{1}{L}
\arcsin(r)<\frac{\pi}{2L}.
\label{rbound1}
\end{equation}
Therefore, for the three-vertex chain graph with $r<1$, the
separating points (\ref{k}) are given by
\begin{equation}
\hat{k}_{n}=\frac{\pi }{L}\left( n+\frac{1}{2}\right).
\label{kchain}
\end{equation}
They mark the borders of the root
intervals $I_n$ for all possible values of the
bond lengths.
For a given prime periodic orbit $p$,
such as the ones shown in Fig.~2, the
coefficients $A_{p}$ are given by \cite{Kar,us}
\begin{equation}
A_{p}= (-1)^{\chi(p)}r^{\sigma(p)}(1-r^{2})^{\tau(p)/2},
\label{A}
\end{equation}
where $r$ is the reflection
coefficient at the middle vertex and $\sigma(p)$
and $\tau(p)$ are correspondingly
the number of reflections off it, and
transmissions through it.
Since the reflection coefficient
may be positive or negative depending on
whether the particle scatters from
the right or from the left, the factor
$(-1)^{\chi(p)}$ is needed to keep
track of how many times it appears with a
minus sign.
Moreover $\chi$ keeps track of how
many times the particle scatters off the
walls of the potential, since each
scattering event from the walls gives rise
to a minus sign in the wave function.
Using the explicit form of the
spectral determinant (\ref{3chain}), one
obtains the explicit form of the
eigenvalues for a quantum particle in the
step potential in terms of classical
(Newtonian and non-Newtonian \cite{Koch}) periodic
orbits.

In order to illustrate the
validity of (\ref{kn}) we computed $k_1$,
$k_{10}$ and $k_{100}$ including
periodic orbits that experience up to
five scattering events.
We chose $b=0.3$ (see Fig. 2)
and $\lambda=1/2$ (see (\ref{scal})).
We obtain $k_1^{(5)}=4.1161$, $k_{10}^{(5)}=39.2866$ and
$k_{100}^{(5)}=394.9477$.
This can be compared with the
exact $k$ values given by $k_1=4.107149$,
$k_{10}=39.305209$ and $k_{100}=394.964713$.
In order to illustrate convergence, we also computed
$k_1$, $k_{10}$ and $k_{100}$
including all periodic orbits that experience up to 20
scattering events. This amounts to including more than
100,000 periodic orbits and results in $k_1^{(20)}=4.105130$,
$k_{10}^{(20)}=39.305212$ and $k_{100}^{(20)}=394.964555$.
This also indicates that the convergence of (\ref{kn})
is not destroyed by keeping more and longer periodic orbits.

Additional examples of regular graphs
are provided by the scaling ``Manhattan potentials''.
These potentials are
obvious generalizations of the step potential shown in
Fig.~2 (a) to piecewise constant potentials where
the potential heights scale with the energy.
Furthermore
we checked explicitly that
linear chain graphs with scaling $\delta$ function potentials
at the vertices provide more examples of regular quantum graphs.
In this case the strengths of the $\delta$ function potentials  scale
linearly with the momentum.

The key physical feature of these
one-dimensional quantum systems, which
permits the exact periodic orbit
expansion (\ref{kn}) for the eigenvalues,
is the rigidity of their spectra.
For integrable systems, spectral
rigidity is due to the ``geometrical
rigidity'' of the periodic orbits,
confined to integrable tori.
In the case of quantum graphs, the
geometrical structure of the periodic
orbits is much more complicated.
The complexity of the expansion (\ref{kn})
compared to the EBK formula
reflects the geometrical complexity of the
periodic orbits.

This paper profitted from a fruitful
exchange of E-mails with Uzy Smilansky.
Y.D. and R.B. gratefully acknowledge financial
support by NSF grants No.
PHY-9900730 and PHY-9984075; Y.D. and R.J by NSF
grant No. PHY-9900746.
%%%%%%%%%%%%%%%%%%%%%%%%%%%%%%%%%%%%%%%%%%%%%%%%%%%%%%%%%%%%%%%%

%\end{multicols}

\pagebreak

\centerline{\bf Figure Captions}

\bigskip \noindent
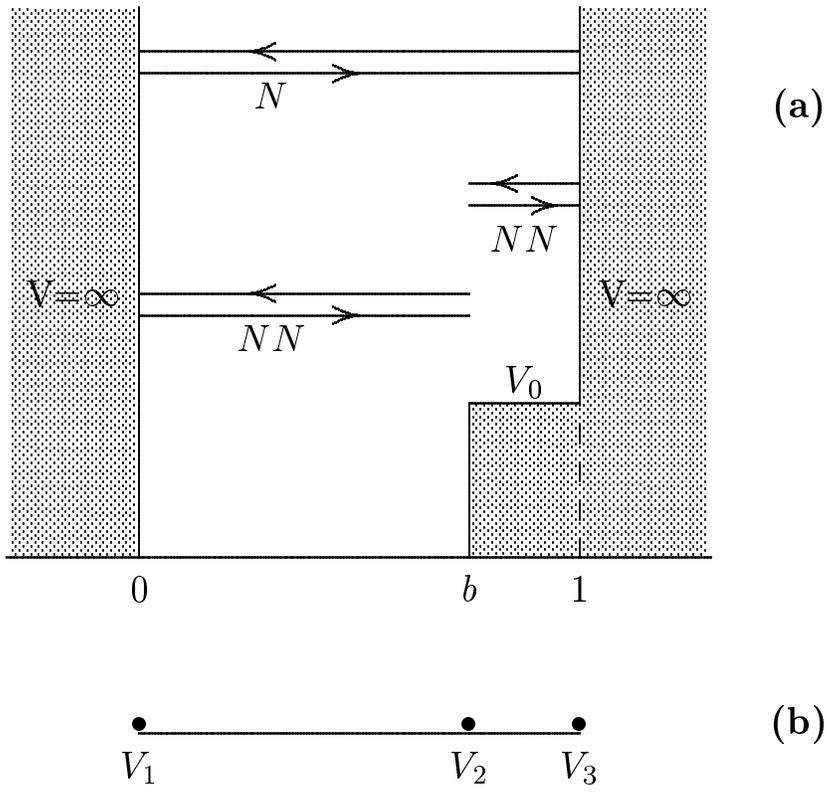
{\bf Fig.~1:} A generic (quantum) graph with five vertices and
  six bonds.

\bigskip \noindent
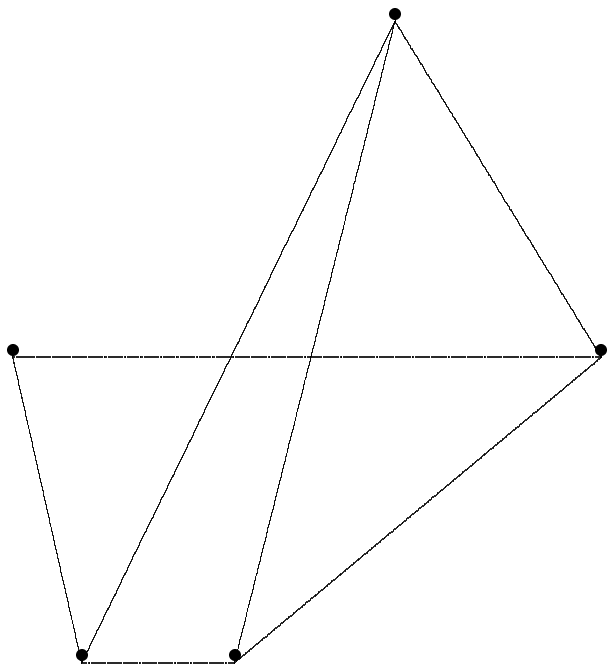
{\bf Fig.~2:} (a) Simple step potential, a basic problem in
  one-dimensional quantum mechanics.
  Also shown are examples of Newtonian (``N'') and
  non-Newtonian (``NN'') periodic orbits used in
  the periodic orbit expansion of its
  energy eigenvalues (see Text). (b)
  Three-vertex chain graph corresponding to
  the step potential above.

\end{document}